\documentstyle[aps]{revtex}

\begin{document}
\title{Improving the capacity of the ping-pong protocol}
\author{Qing-yu Cai and Bai-wen Li}
\address{Wuhan institute of Physics and Mathematics, The Chinese Academy of Science,\\
Wuhan, 430071, The People's Republic of China}
\maketitle

\begin{abstract}
We present a quantum communication protocol which keeps all the properties
of the ping-pong protocol [Phys. Rev. Lett. 89, 187902 (2002)] but improves
the capacity doubly as the ping-pong protocol. Alice and Bob can use the
variable measurement basises in control mode to detect Eve's eavesdropping
attack. In message mode, Alice can use one unitary operations to encode two
bits information. Bob only needs to perform a Bell type measurement to
decode Alice's information. A classical message authentification method can
protect this protocol against the eavesdropping hiding in the quantum
channel losses and the denial-of-service (DoS) attack.
\end{abstract}


Since Bennett and Brassard presented the pioneer quantum key distribution
(QKD) protocol in 1984 [1], there are a lot of theoretical QKD schemes
[1-12] today. Recently, Beige $et$ $al$. proposed a quantum secure direct
communication scheme [13], which allows messages communicated directly
without first establishing a random key to encrypt them. Bostroem and
Felbingeer presented a ping-pong protocol [14] which is secure for key
distribution and quasisecure for direct communication with a perfect quantum
channel. However, an eavesdropper-Eve can eavesdrop the information if the
quantum channel is noisy [15]. And this ping-pong protocol can attacked
without eavesdropping [16]. Moreover, the capacity is restricted, and an
entangled state only carries one bit of classical information in every
message run. In this paper, we give an improved ping-pong protocol, which
allows one entangled state carrying two classical bits in every message run.
And we show this protocol is secure with a classical message
authentification method.

Let us start with the brief description of the ping-pong protocol. Suppose
Bob have two photons that are maximally entangled in their polarization
degree of freedom, $|\psi ^{\pm }>=\frac{1}{\sqrt{2}}(|1>|0>\pm |0>|1>)$.
Since the reduced density matrices, $\rho _{A}=tr_{B}\{|\psi ^{\pm }><\psi
^{\pm }|\}$ are completely mixture, then one can not distinguish the state $%
|\psi ^{\pm }>$ from each other by only access one qubit. Since the state $%
|\psi ^{\pm }>$ are mutually orthogonal , a measurement on both qubit can
perfectly distinguish the states form each other. Bob sends one of the
photon (travel photon) to Alice and keeps another (home qubit). Alice can
performs a unitary operation $U_{j}$, 
\begin{equation}
U_{j}=(|0><0|-|1><1|)^{j},
\end{equation}
where $j\in \{0,1\}$, on the travel qubit to encode her information $j$.
Then she sends this qubit back to Bob. When Bob receives the travel qubit,
she performs a joint measurement on both photons to decode Alice's
information. To confirm the security of this communication, Alice randomly
switches the message mode to control mode. In control mode Alice measures
the travel qubit in basis $B_{z}=\{|0>,|1>\}$and announces the result in the
public channel. When receiving Alice's result, Bob also switches to control
mode. Bob measures the home qubit in basis $B_{z}$ and compares the both
results. If both results coincide, there is Eve in line. Else, this
communication continues.

In every message run, Alice encodes one bit on the travel qubit. As it is
well known, there are four Bell states that are mutual orthogonal to each
other, $|\psi ^{\pm }>$, $|\phi ^{\pm }>$. And local unitary operation can
transform these four states to each other. The Bell's states can be written
as 
\begin{eqnarray}
|\phi ^{+} &>&=\frac{1}{\sqrt{2}}(|0>|0>+|1>|1>)  \nonumber \\
&=&\frac{1}{\sqrt{2}}(|+>|+>+|->|->),
\end{eqnarray}
\begin{eqnarray}
|\phi ^{-} &>&=\frac{1}{\sqrt{2}}(|0>|0>-|1>|1>)  \nonumber \\
&=&\frac{1}{\sqrt{2}}(|+>|->+|->|+>),
\end{eqnarray}
\begin{eqnarray}
|\psi ^{+} &>&=\frac{1}{\sqrt{2}}(|0>|1>+|1>|0>)  \nonumber \\
&=&\frac{1}{\sqrt{2}}(|+>|+>-|->|->),
\end{eqnarray}
\begin{eqnarray}
|\psi ^{+} &>&=\frac{1}{\sqrt{2}}(|0>|1>-|1>|0>)  \nonumber \\
&=&\frac{1}{\sqrt{2}}(|+>|->-|->|+>),
\end{eqnarray}
where $|+>=\frac{1}{\sqrt{2}}(|0>+|1>)$, $|->=\frac{1}{\sqrt{2}}(|0>-|1>)$.
Suppose Bob prepare an EPR pair in state $|\psi ^{-}>$. In message mode,
Alice performs a unitary $U_{ij}$ to encode her information, where $U_{ij}$
are 
\begin{eqnarray}
U_{00} &=&\left( 
\begin{array}{ll}
1 & 0 \\ 
0 & 1
\end{array}
\right) ,U_{01}=\left( 
\begin{array}{ll}
1 & 0 \\ 
0 & -1
\end{array}
\right) ,  \nonumber \\
U_{10} &=&\left( 
\begin{array}{ll}
0 & 1 \\ 
1 & 0
\end{array}
\right) ,U_{11}=\left( 
\begin{array}{ll}
0 & 1 \\ 
-1 & 0
\end{array}
\right) ,
\end{eqnarray}
corresponding to Alice's 00,01,10,11. When Bob receives the travel back
qubit, he can performs a Bell type measurement in basis $\{|\psi ^{\pm
}>,|\phi ^{\pm }>\}$ to decode Alice's information. Then one EPR pair can
transmit two bits in every message run.

To ensure the security of this protocol, the control mode should be
modified. In control mode. Alice receives the travel qubit. Alice performs a
measurement randomly in the basis $B_{z}=\{|0>,|1>\}$ or $B_{x}=\{|+>,|->\}$%
. Then she announces her measurement result and the basis through public
channel. Bob also switches to control mode. He performs a measurement in the
same basis Alice used. If both results coincide, then there is Eve in line.
Else, Bob sends next qubit to Alice.

Since $\rho _{A}=tr_{B}\{|\psi ^{\pm }><\psi ^{\pm }|\}=tr_{B}\{|\phi ^{\pm
}><^{\pm }\phi |\}=\frac{1}{2}|0><0|+\frac{1}{2}|1><1|$, Eve can not
distinguish each Bell state if she only attack the travel after Alice
encoding operation. So she has to attack the travel in line $B\rightarrow A$
first. After Alice's encoding operation, she performs a measurement attack
in line $A\rightarrow B$ to draw Alice's information. Let us assume that the
state $|\psi ^{-}>$ becomes $\rho $ after Eve's attack in line $B\rightarrow
A$. Then the information Eve can gain from $\rho $ is bounded by the Holevo
quantity, $\chi (\rho )$ [17]. Because Holevo quantity decreases under
quantum operations [17,18], then the mutual information Eve can gain after
Alice's encoding operation is determined by $\chi (\rho )$. From 
\begin{equation}
\chi (\rho )=S(\rho )-\sum_{i}p_{i}S(\rho _{i}),
\end{equation}
we know $S(\rho )$ is the upper bound of $\chi (\rho )$. `High fidelity
implies low entropy'. Suppose 
\begin{equation}
F(|\psi ^{-}>,\rho )^{2}=<\psi ^{-}|\rho |\psi ^{-}>=1-\gamma ,
\end{equation}
where $F(|\psi ^{-}>,\rho )$ is the fidelity [19] of state $|\psi ^{-}>$ and 
$\rho $, $0\leq \gamma \leq 1$. Therefore, the entropy of $\rho $ is bounded
above by the entropy of a diagonal density matrix $\rho _{\max }$ with
diagonal entries $1-\gamma $, $\frac{\gamma }{3}$, $\frac{\gamma }{3}$, $%
\frac{\gamma }{3}$. Then entropy of $\rho _{\max }$ is 
\begin{equation}
S(\rho _{\max })=-(1-\gamma )\log (1-\gamma )-\gamma \log \frac{\gamma }{3}.
\end{equation}
Let us discuss the relation between the fidelity $F(|\psi ^{-}>,\rho )$ and
the detection probability $d$. In control mode, when the states Alice and
Bob shared are $|\phi ^{\pm }>$, their measurement results will coincide
every time when they use the measurement $B_{z}$. When the state they shared
is $|\psi ^{+}>$, their measurement results will coincide when they use the
measurement basis $B_{x}$. Only their state is $|\psi ^{-}>$, they
measurement results will never coincide. Since $F(|\psi ^{-}>,\rho
)^{2}=1-\gamma $, then the detection probability is $d\geq \gamma /2$. From
Eq.(9), we know when $\gamma =0$, i.e., Eve does not attack the state $|\psi
^{-}>$ in line $B\rightarrow A$, the detection probability $d=0$. When $%
\gamma >0$, i.e., Eve can gain some of Alice's information, she has to face
a nonzero risk $d\geq \gamma /2$ to be detected. When $\gamma =\frac{3}{4}$,
it has $S(\rho _{\max })=2$, which implies Eve has chance to eavesdrop full
of Alice's information. On this condition, it has the maximal detection
probability $d\geq \frac{3}{8}$.

W$\stackrel{^{\prime }}{o}$jcik presented an eavesdropping scheme [15] which
reveals that the ping-pong is not secure for transmission efficiencies lower
than 60\%. A classical message authentification method can protect this
ping-pong protocol against the eavesdropping hiding in the quantum channel
losses. Actually, in this improved ping-pong protocol, the eavesdropping
hiding in quantum channel losses can be detected because of the variable
measurement basises in control mode.

In Ref.[16], it has been shown that the ping-pong protocol can be attacked
without eavesdropping. Eve can attack the travel qubit in line $A\rightarrow
B$. She can perform a unitary operation on the travel qubit to change the
states. Also, she can performs a measurement on the travel qubit to destroy
the EPR states. But any message authentification method can protect this
protocol against the man-in-the-middle attacks with a reliable public
channel.

In summary, we present a protocol which keeps all properties of the
ping-pong protocol but improves the capacity doubly as the ping-pong
protocol. And this protocol can protect the eavesdropping hiding in quantum
channel losses. And the message authentification method can protect the
protocol against the man-in-the-middle attacks.

\section{references}

[1] C. H. Bennett, and G. Brassard, in Proceedings of the IEEE international
Conference on Computers, Systems and Signal Processing, Bangalore, India
(IEEE, New York, 1984), pp. 175-179.

[2] A. K. Ekert, Phys. Rev. Lett. 67, 661 (1991).

[3] C. H. Bennett, G. Brassard, and N. D. Mermin, Phys. Rev. Lett. 68, 557
(1992).

[4] C. H. Bennett, Phys. Rev. Lett. 68, 3121 (1992).

[5] C. H. Bennett and S. J. Winsner, Phys. Rev. Lett. 69, 2881 (1992).

[6] L. Goldenberg and L. Vaidman, Phys. Rev. Lett. 75, 1239 (1995).

[7] B. Huttner, N. Imoto, N. Gisin, and T. Mor, Phys. Rev. A 51, 1863 (1995).

[8] M. Koashi and N. Imoto, Phys. Rev. Lett. 79, 2383 (19970.

[9] D. Bruss, Phys. Rev. Lett. 81, 3018 (1998).

[10] W. Y. Hwang, I. G. Koh, and Y. D. Han, Phys. Lett. A 244, 489 (1998).

[11] H.-K. Lo and H. F. Chau, Science 283, 2050 (1999).

[12] A. Cabell, Phys. Rev. Lett. 85, 5635 (2000).

[13] A. Beige, B.-G. Englert, C. Kurtsiefer, and H. Weinfurter, Acta. Phys.
Pol. A 101, 357 (2002).

[14] K. Bostroem and T. Felbinger, Phys. Rev. Lett. 89, 187902 (2002).

[15] A. W$\stackrel{^{\prime }}{o}$jcik, Phys. Rev. Lett. 90, 157901 (2003).

[16] Qing-yu Cai, Phys. Rev. Lett. 91, 109801 (2003).

[17] M. A. Nielsen and I. L. Chuang, Quantum computation and Quantum
Information (Cambridge University Press, Cambridge, UK, 2000).

[18] Qing-yu Cai, arXiv:quant-ph/0303117(unpublished).

[19] C. A. Fuchs, arXiv: quant-ph/9601020; H. Barnum, C. M. Caves, C. A.
Fuchs, R. Jozsa, and B. Schumacher, Phys. Rev. Lett. 76, 2818 (1996).

\end{document}